\documentclass{elsart}
\usepackage{amsfonts,amsmath,graphicx,hyperref}
\usepackage[numbers]{natbib}
\usepackage{lineno}

\makeatletter
\newif\ifwithtwocolumns\withtwocolumnsfalse
\if@TwoColumn\withtwocolumnstrue\fi
\makeatother

\newcommand{\tmem}[1]{{\em #1\/}}
\newcommand{\tmop}[1]{\ensuremath{\operatorname{#1}}}

\newcommand{\tmtexttt}[1]{{\ttfamily{#1}}}

\def\hc#1{\leavevmode\hbox to \hsize{\hss #1\hss}\leavevmode}

\begin{document}


\begin{frontmatter}
  \title{Simulation of Imaging Atmospheric Cherenkov Telescopes with
    CORSIKA and \tmtexttt{sim\_telarray}}
  
  \author[mpik,hub]{Konrad Bernl\"ohr}
  \ead{Konrad.Bernloehr@mpi-hd.mpg.de}
  \address[mpik]{Max-Planck-Institut f\"ur Kernphysik, Postfach 103980, 69029 Heidelberg, Germany}
  \address[hub]{Institut f{\"u}r Physik, Humboldt-Universit\"at zu Berlin, Newtonstra{\ss}e 15, 12489 Berlin, Germany}
  
  \begin{abstract}
    Imaging Atmospheric Cherenkov Telescopes (IACTs) have resulted in a
    break-through in very-high energy (VHE) gamma-ray astrophysics. While
    early IACT installations faced the problem of detecting any sources at
    all, current instruments are able to see many sources, often over more
    than two orders of magnitude in energy. As instruments and analysis
    methods have matured, the requirements for calibration and modelling of
    physical and instrumental effects have increased. In this article, a set
    of Monte Carlo simulation tools is described that attempts to include all
    relevant effects for IACTs in great detail but aims to achieve this in an
    efficient and flexible way. These tools were originally developed for the
    HEGRA IACT system and later adapted for the H.E.S.S. experiment. Their
    inherent flexibility to describe quite arbitrary IACT systems makes them
    also an ideal tool for evaluating the potential of future installations.
    It is in use for design studies of CTA and other projects.
  \end{abstract}

\end{frontmatter}



\section{Introduction}

Any simulation of the IACT technique consists of two major components: the
development of the extensive air showers and emission of Cherenkov light by
the shower particles as one part and the detection of this light and recording
of signals by the instrument as the other part. While earlier simulation tools
used to come with a rather primitive description of the instrumental response
included with the shower simulation, the simulation tools described here
separates these parts into distinct programmes. This allows for enhanced
flexibility in the implementation and sharing of components in the community.

The first major component is based on the CORSIKA program
{\cite{CORSIKA-physics}}. A first implementation of Cherenkov light in
CORSIKA, by M.~Rozanska, F.~Arqueros, and S.~Martinez, implemented a
rectangular grid of detectors -- for the non-imaging HEGRA AIROBICC instrument
{\cite{1995NIMPA.357..567M}}. The current implementation, by the author of
this paper and with contributions by others, unified the originally separate
functions for Cherenkov emission by electrons/positrons and by other
particles, improved its efficiency and accuracy, and added a great level of
flexibility. The CORSIKA code base includes different output formats now, some
more generic and some dedicated to specific experiments. Only the most generic
of these, the IACT option also developed by us, will be covered here. With the
IACT option, a telescope or other Cherenkov detector in CORSIKA is defined by
a fiducial sphere containing the reflector or detector.

\begin{figure*}[tb]
\ifwithtwocolumns
  \hc{\resizebox{0.8\textwidth}{!}{\includegraphics{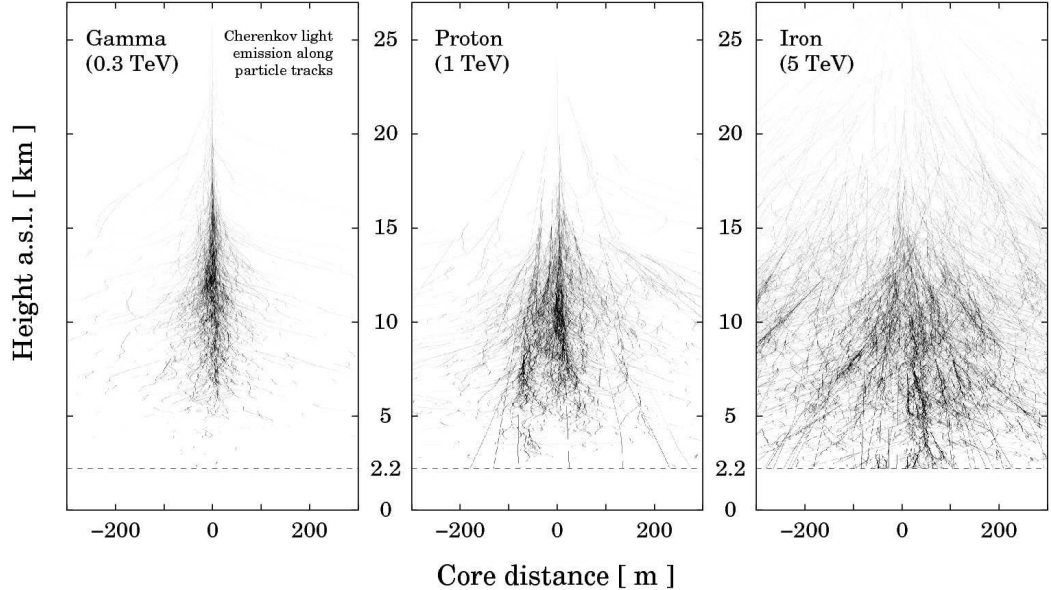}}}
\else
  \resizebox{13.8cm}{!}{\includegraphics{gamma+p+iron_n}}
\fi
  \caption{Cherenkov light production in CORSIKA for different primary
  particle examples in simulations for a site at 2200~m altitude. Darkness of
  the particle tracks shown increases with increasing emission of Cherenkov
  light.\label{fig:shower-primaries}}
\end{figure*}

The second major component, termed \tmtexttt{sim\_tel\-array} or -- in its
current incarnation for the H.E.S.S. experiment \citep{HESS-LOI} -- \tmtexttt{sim\_hess\-array},
implements all the details of the detectors. This includes optical ray-tracing
of the photons, their registration by photomultiplier tubes, switching of
discriminators or comparators at the pixel level and at the telescope trigger
level, as well as digitization of the resulting signals. Since
\tmtexttt{sim\_telarray} is highly efficient and, in a typical configuration,
requires only a small fraction of the CPU time needed for the shower
simulation, it is usually run several times in parallel, reading directly from
a common output pipe of the CORSIKA IACT option. Each instance might
correspond to a different viewing direction, a different atmospheric
transmission, or even a completely different mirror and camera definition.


\section{CORSIKA and the IACT option}

\subsection{Cherenkov light production in CORSIKA}

Cherenkov light production in CORSIKA (see Figure \ref{fig:shower-primaries}) 
with its various options\footnote{CORSIKA options discussed in this paper
include:

 ATMEXT: atmospheric extension, providing tabulated atmospheric
    profiles to all CORSIKA variants and refraction to the Cherenkov light.

 CEFFIC: Very simple treatment of atmospheric transmission, mirror
    reflectivity, and quantum efficiency in CORSIKA itself.

 CERENKOV: the master switch for enabling Cherenkov emission.
    Without IACT option, a rectangular detector array is simulated.

 CERWLEN: the index of refraction and, thus, the emission angle
    depends on the wavelength.

 IACT: the detectors are defined individually as fiducial spheres
    and the machine-independent output data format readable by 
    \tmtexttt{sim\_telarray} gets used.
}
(like IACT, ATMEXT, CEFFIC, CERWLEN, ...) has to be enabled before
compilation (see \citep{CORSIKA-user} for details). 
This results in a call to the Cherenkov subroutine CERENK for each track
segment of charged particles as they are handled by the CORSIKA particle
transport code. The transport code takes care of possible interactions, decay,
multiple scattering, bending in the geomagnetic field, and ionization energy
loss. As far as Cherenkov light production is concerned, each straight track
segment is defined by start and end point as well as by the initial and final
energy of the particle (assuming continuous energy loss), and its mass and
charge.

Since the track segments can have lengths up to several kilometres at high
altitudes, the Cherenkov emission subroutine in CORSIKA has to take 
care of changes in the index of
refraction of the air as well as of the change in velocity of the particle.
Track segments in CORSIKA are generally shorter when Cherenkov light
production is enabled, in particular with the IACT option, where typical
multiple scattering angles and bending in the geomagnetic field should be
smaller than the pixel scale in the cameras. Too long track segments
could result, for example, in too sharp muon rings or even a shift in
the shower maximum. It has been verified that the
reduced step lengths with the IACT option (compared to those used
in CORSIKA without Cherenkov light production and STEPFC=1.0) do not change 
the amount of Cherenkov light or its average lateral distribution in any 
noticeable way, i.e.\ well below 1 percent.

\begin{figure}[tb]
\ifwithtwocolumns
  \hc{\resizebox{\hsize}{!}{\includegraphics{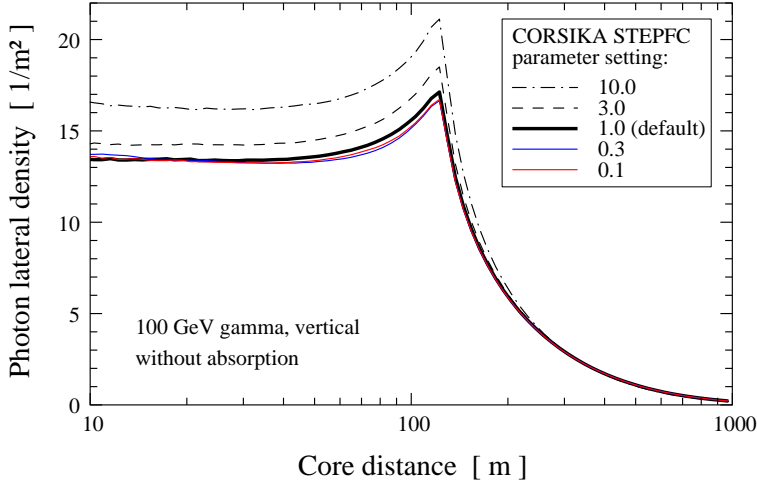}}}
\else
  \resizebox{10cm}{!}{\includegraphics{lat_6_stepfc}}
\fi
  \caption{Average lateral distribution of 300-600 nm Cherenkov light in simulations
  of vertical 100 GeV gamma-rays for a site at 2200~m altitude, with different
  settings of the EGS step length parameter STEPFC. Too long step lengths result
  in a systematic excess of light inside the {\em light pool} ($r<120$~m).
  \label{fig:stepfc}}
\end{figure}

The STEPFC parameter of CORSIKA could be used to change step lengths
of gammas and $e^\pm$ relative to the default settings in EGS
{\citep{EGS}}, which correspond to STEPFC=1.0. While longer
step lengths could be used
to improve the processing speed, they result in a systematic
excess of Cherenkov light by about 20\% for STEPFC=10 (see
Figure~\ref{fig:stepfc}).
It is reassuring to find that further reductions of step lengths 
(e.g.\ STEPFC=0.1) will not result in another systematic change.
The question of proper step lengths is certainly an important
issue for any shower simulation program -- for CORSIKA the
built-in default appears to be just the right choice.

While accurate modelling of the showers is perhaps the prime objective in
CORSIKA, efficiency is important as well. The efficiency aspect is quite
important in the case of the Cherenkov emission subroutine since most of the CPU time is
spent there. For many applications, the most important performance gain is
achieved by the fact that Cherenkov photons are not simulated one by one but
in {\tmem{bunches}}. This concept of bunches was already present in the
original implementation {\citep{1995NIMPA.357..567M}}, with an automatically
adapted bunch size as a function of the energy of the primary particle,
optimized for the AIROBICC experiment.

\begin{figure*}[tb]
\ifwithtwocolumns
  \hc{\resizebox{0.8\textwidth}{!}{\includegraphics{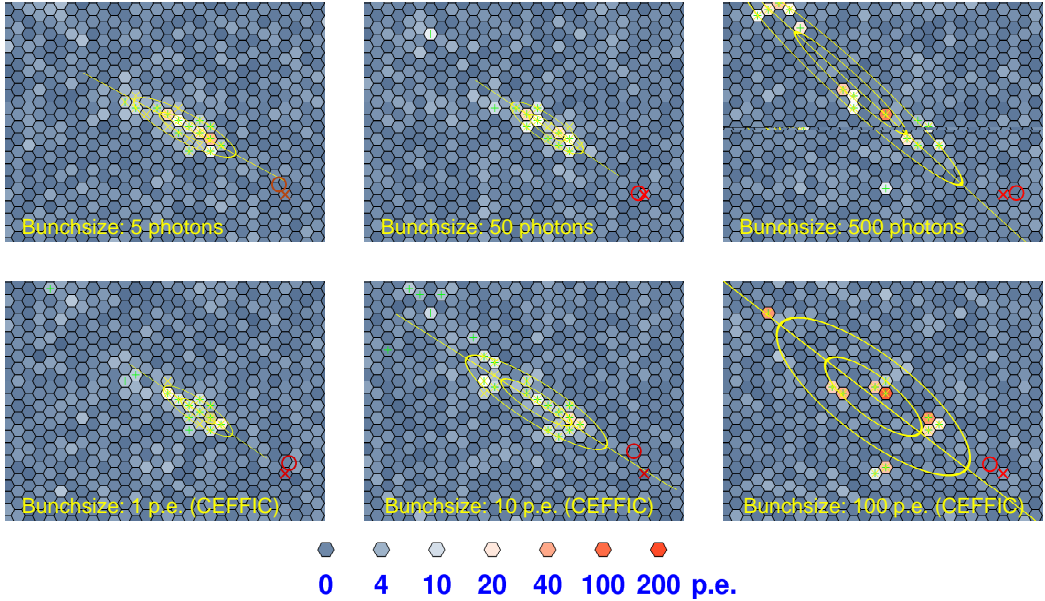}}}
\else
  \resizebox{13.8cm}{!}{\includegraphics{bunchsize}}
\fi
  \caption{The effect of different bunch sizes on the image obtained with simulated IACTs.
  All samples show the same $1.8^\circ\times1.35^\circ$ part of the field-of-view
  of a 24~m telescope looking at a 62~GeV $\gamma$-ray shower from 117~m core distance.
  The red cross in each panel indicates the true direction of the incident $\gamma$-ray
  (for other symbols see also Figure~\ref{fig:event-display}). The colour scale at
  the bottom indicates the intensities found in each 0.07$^\circ$ pixel, in units
  of the average intensity from single photo-electrons (p.e.).
  Top row: No CEFFIC option, atmospheric transmission and photon detection handled
  only by \tmtexttt{sim\_telarray}; 
  left to right: bunch sizes of 5, 50, and 500 (recommended: 5 to 10).
  Bottom row: With CEFFIC option, detection efficiency applied in CORSIKA:
  bunch sizes of 1, 10, and 100 (recommended: $\leq$1). Bunch sizes in the left column are
  small enough to show the actual smooth image, in the middle column artificial fluctuations
  are already apparent, and in the right column the artificial fluctuations dominated over
  the true fluctuations in the shower.
  \label{fig:bunch-size}}
\end{figure*}

For imaging telescopes, an automatic selection of bunch sizes is not so easy.
Showers induced by high-energy primaries can be detected from large distances
while less energetic showers are typically only detected in the region of the
almost uniformly illuminated central {\tmem{Cherenkov light pool}}. The
optimum bunch size is highly dependent on the detector configuration and best
defined by the user in the CORSIKA {\tmem{inputs}} file. See
{\citep{CORSIKA-user}} for instructions. Too low values are always safe, but
inefficient, while too high values will result in artificial image
fluctuations, as illustrated in Figure~\ref{fig:bunch-size}. 
In a typical configuration with conventional PMTs, a bunch size
of 5 to 10 for a wavelength range from 250 to 700 nm should be fairly safe
without the CEFFIC option. With the CEFFIC option, where the detection
efficiency is included in CORSIKA itself instead of the detector simulation,
bunch sizes above one can generally not be recommended.

For a given bunch size, the track segment is sub-divided into steps such that
in each step one photon bunch is emitted into a random direction on the
Cherenkov cone. The actual size of each bunch and the Cherenkov cone opening
angle, defined by $\cos \theta = 1 / (\beta n)$, depend on the index of
refraction $n$ and the velocity $v = \beta c$ of the particle at the emission
point. The wavelength $\lambda$ of the photons is usually unspecified (unless the
CEFFIC or CERWLEN options are used) and may
be randomized by the detector simulation to follow the $1 / \lambda^2$
distribution in the predefined wavelength range.

\begin{figure}[htb]
\ifwithtwocolumns
  \resizebox{\hsize}{!}{\includegraphics{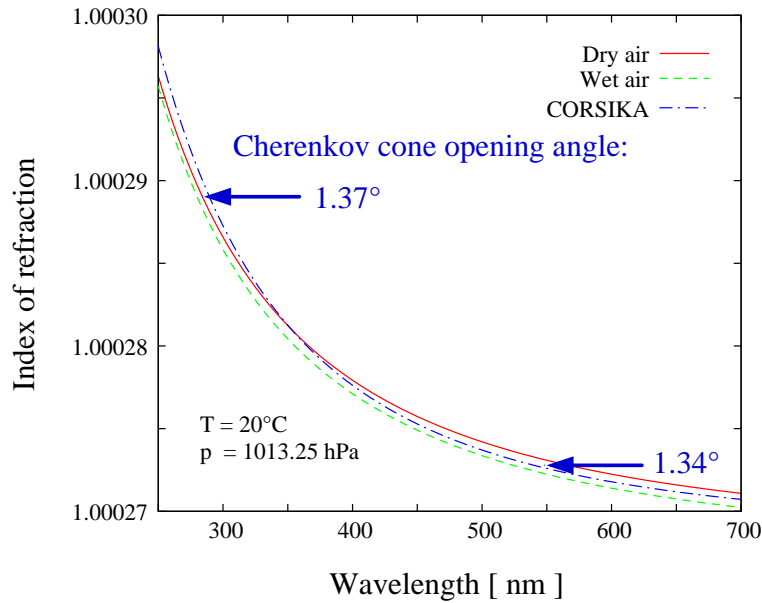}}
\else
  \resizebox{10cm}{!}{\includegraphics{refidx3}}
\fi
  \caption{Wavelength dependence of the index of refraction 
  \citep{refractive-index} and its approximation in CORSIKA with
  the CERWLEN option.
  The index of refraction as used in CORSIKA typically corresponds to a
  wavelength of 400~nm. 
  Also shown is the Cherenkov
  emission angle for a very fast particle under normal sea-level conditions.
  Maximum opening angles are 1.37$^{\circ}$ and
  1.34$^{\circ}$ for 300 and 550 nm, respectively,
  \label{fig:refraction-wl}under the given conditions.}
\end{figure}

Where even more detailed simulations are desired, a random wavelength can be
selected in CORSIKA and the wavelength dependence of the index of refraction
(see Figure \ref{fig:refraction-wl}) taken into account to determine the
emission angle and the amount of light, i.e.\ bunch size. This level of detail
is not normally needed and has a significant impact on performance, and is
therefore only enabled on request (CERWLEN option). Normally, the index of
refraction $n$ is assumed independent of wavelength and should correspond to
an effective wavelength, by default 400~nm. Depending on the configured
atmospheric profile, the refractivity $n - 1$ is assumed proportional to the
density of the air or interpolated from a table where the dependence on
humidity and temperature may be included, typically based on radiosonde data
for the site in question.

\subsection{The IACT/ATMO package}

The original Cherenkov light implementation in CORSIKA itself was designed to
match a rectangular detector array on a horizontal plane. This doesn't allow
to match most actual or non-horizontal detector layouts. For large
zenith-angle observations, the concept of horizontally flat detectors will
even have difficulties with rectangular telescope arrays.

An extension package for CORSIKA, written in the C language, overcomes these
limitations. The IACT/ATMO package, together with its corresponding interfaces
in CORSIKA, serves several purposes: the geometry of an IACT array can be matched
to actual configurations, use of a machine- and compiler-independent data
format, use of tabulated atmospheric profiles, and taking atmospheric
refraction into account.

An {\tmem{array}} of telescopes or other detectors is defined by the $(x_i,
y_i, z_i)$ positions of the centres and the radii $r_i$ of any number of
fiducial spheres (currently limited to 1000 in CORSIKA input handling). 
For a typical IACT the position would best correspond to the
intersection of altitude and azimuth axes and the radius would be large enough
to enclose the whole reflector. No other properties of the instrument are
required in this part of the simulation. For efficient use of the CPU time
invested in shower and Cherenkov light simulation, the defined array can be
re-used multiple times for each shower at random displacements with respect to
the nominal shower core. By default, a circle of given radius $R_{c, \max}$
will be uniformly covered. Depending on the CORSIKA compile-time configuration
this can either be a circle in the horizontal {\tmem{detection plane}} as used
in CORSIKA without IACT option or in the {\tmem{shower plane}} perpendicular
to the direction of the primary particle and projecting them back to the
detection level. 

Non-uniform coverage is optionally available for 
implementation by the user. This could be used, for example, for generating 
low energy showers over a smaller core distance range than high energy showers. 
Event weights, which are then needed for the reconstruction of
spectra, are kept in the provided data format. In the more general case,
the non-uniform coverage can support variance reduction schemes for
specific questions, i.e.\ reducing the variance in some resulting number
(like the trigger rate, for example) for a given number of showers --
or reducing the number of showers required for a given variance.
Usually, preference has to be given to the core positions more likely resulting
in a trigger, in order to achieve that (thus often called 
{\em importance sampling} \citep{Rosenbaum1971}) -- but the optimum
distribution depends on the question and detector details and cannot be generalised.

The nominal core position in CORSIKA is along a straight line on the original
direction of the primary particle from its starting position at the assumed
top of the atmosphere -- typically 100 to 120 km high. Low-energy charged
particles, e.g.\ electrons of a few GeV, suffer substantial bending in the
geomagnetic field before their first interaction. For electrons detectable
with large IACTs the deflection can approach $R_{c, \max}$ and would affect
detection probabilities even under assumption of an isotropic background. The
deflection is therefore compensated in the IACT/ATMO package.

\begin{figure}[htb]
\ifwithtwocolumns
  \resizebox{\hsize}{!}{\includegraphics{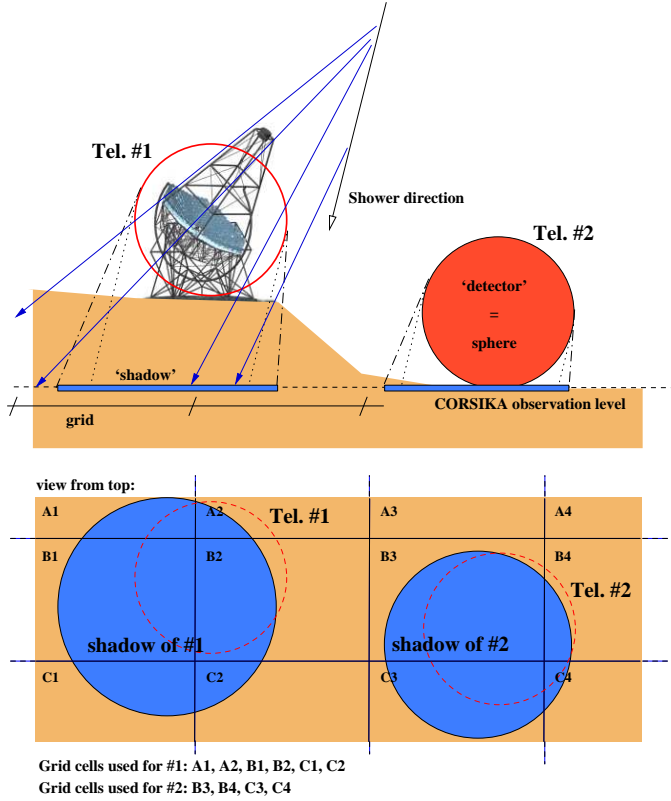}}
\else
  \resizebox{9cm}{!}{\includegraphics{iact3d-nt}}
\fi
  \caption{Definition of grid cells at the detection level to which a detector
  sphere gets linked for detailed inspection of intersections of photon
  bunches with the sphere. The {\tmem{shadow}} of a sphere is large enough to
  include all Cherenkov light emitted up to 10 degrees from the shower
  direction and intersecting the sphere. For detectors near the
  CORSIKA observation level even most light emitted at larger angles
  would be recorded. Some rays at extremely large angles
  not relevant for IACTs (but perhaps for fluorescence detectors) could go 
  unnoticed for efficiency reasons.\label{fig:iact3d}}
\end{figure}

Since the possible intersection of each photon path with many spheres would
be very CPU intense, a rectangular grid at the detection level is set up and
each sphere is linked to one or more of the grid cells as illustrated in
Figure~\ref{fig:iact3d}. Only spheres linked to the grid cell, where a photon
bunch arrives, are checked for possible intersections. This scheme is very
effective and the CPU time required for accumulating photon bunches hitting
any detector is generally negligible. In order to reduce memory requirements,
a thinning scheme can set in when too many bunches get collected for a
telescope, allowing for high dynamic range simulations. 
In this thinning scheme, similar to thinning of particles in CORSIKA, the
number of bunches is reduced but the number of photons in each of the
remaining bunches is increased in compensation.

The photon bunches intersecting any of the detector spheres in any of the
randomly displaced array instances are recorded in a machine- and
compiler-independent, flexible data format, termed {\tmem{eventio}}, which was
originally developed for other purposes {\citep{CRT,HEGRA}}.
Photon bunches in this format can have an unspecified or a specific
wavelength, with negative wavelengths indicating that detection efficiencies
were already applied (CEFFIC option, bunches corresponding to
photo-electrons). The output can be written to file, with optional automatic
compression, or it can be piped into another program, either the telescope
simulation directly or into an adapter program which in turn pipes the output
into several different telescope simulations. This way, different telescope
configurations, different offset angles of the telescopes with respect to the
source, and different atmospheric transmission models can be simulated at the
same time. Hundreds of shower simulations, with thousands of telescope
simulations, can be processed in parallel on a large computer cluster without
running into any I/O bottlenecks.%
\footnote{With little more than 10 telescope simulations reading CORSIKA
data from the same network-mounted RAID array of hard disks, the I/O bandwidth
would be saturated. Piping CORSIKA data directly into \tmtexttt{sim\_telarray},
hundreds of computer nodes may be running 10 telescope simulations each without
running into I/O limitations.}

Another aspect of the IACT/ATMO package, supporting the CORSIKA option
ATMEXT, is the possibility to provide
tabulated atmospheric profiles. These include the density, the atmospheric
thickness, and the effective index of refraction as a function of altitude.
Such a table would typically be derived from radiosonde data for the site in
question or from atmospheric models. Since the EGS4-based part of CORSIKA
cannot interpolate from this table but makes use of several atmospheric layers
with piece-wise exponential or (in the top layer) a linear density profile,
the corresponding parameters are fitted simultaneously to the atmospheric
thickness and density columns at program start-up. Where suitable, the table values are
interpolated by taking advantage of the close to exponential density profile.

Since systems of IACTs can achieve an accuracy on the level of a few arc
seconds in the positions of point-like sources, atmospheric refraction of
Cherenkov light has to be taken into account as well
{\citep{Bernloehr-atmospheric}}. The refraction correction is provided by the
IACT/ATMO package when the CORSIKA option ATMEXT is enabled. 
It affects the arrival direction, the arrival position, and the
arrival time at the CORSIKA detection level. These corrections are evaluated
once by numerical integration and then interpolated as functions of starting
altitude and zenith angle. Fortunately, the differential refraction, i.e.\ the
dependence on wavelength, is not relevant yet for the current systems of
IACTs.

Scattering of Cherenkov light is optionally available, 
including both Rayleigh and Mie scattering. 
However, this requires preparation of special data files, 
is very CPU-intense, and thus not normally used.
In addition, scattered Cherenkov light is irrelevant within the short
integration times of IACTs {\citep{Bernloehr-atmospheric}}.
The scattering option is therefore not included in the package
version distributed with CORSIKA but available on request only.


\section{The \tmtexttt{sim\_telarray} telescope simulation package}

\subsection{Overview}

The \tmtexttt{sim\_telarray} package was originally developed for the HEGRA
IACT system {\citep{HEGRA}}. In this early implementation, many aspects were still hard-wired,
e.g.\ that the telescope trigger would be a next-neighbour trigger with a pixel
logic corresponding to discriminators with immediate switching and sharp
rising and falling edges of the signal. For the design issues with H.E.S.S.
and future telescope systems many aspects were eventually generalized and are
fully configurable at run-time. Each telescope can be separately configured,
including its optical layout and properties as well as the camera
configuration, how telescope and system trigger conditions are formed, and
also how the signals get digitized and recorded. In its current
implementation, it is also referred to as the \tmtexttt{sim\_hessarray}
package although it is in no way specific to H.E.S.S. By exchanging a number
of configuration files, any other IACT system can be simulated just as well.

Realistic and detailed detector simulations are key features of the
\tmtexttt{sim\_telarray} package but efficiency is just as important. Often,
simulation studies require a billion and more events, sometimes with many
telescopes in each event -- in some studies for CTA {\citep{CTA-MC-ICRC}}
close to 100. Full simulation of night-sky background (NSB) and electronic
noise in every pixel of every telescope event would be prohibitively slow for
such studies. In other cases, e.g.\ for finding reasonable trigger conditions
where random triggers due to night-sky background are at a manageable level,
the full simulation of each and every event is unavoidable.

As long as night-sky induced triggers are negligible and most simulated
events will not trigger a telescope, \tmtexttt{sim\_telarray} offers several
short-cuts in the simulation which can boost efficiency by large amounts
without losing events that would eventually trigger a telescope. These
short-cuts require initial full simulations to determine safe lower limits for
the numbers of Cherenkov photons and the resulting number of photo-electrons
needed to trigger a telescope (except for NSB-induced triggers). Where the
required number of photons -- based on the sum of all photon bunches hitting
the fiducial sphere -- is not met, the simulation can be bypassed completely.
Otherwise the optical ray-tracing and photon detection are carried out and the
sum of all photo-electrons is compared to the second limit. Only where that is
met as well we have to go through the detailed simulation of the electronics
response. The difference between complete bypassing and full simulation without
much Cherenkov light can be up to 5 orders of magnitude in time, with little 
more than 0.01 seconds for the full simulation of a H.E.S.S.-I type telescope 
on a fast CPU core. In real productions, the short-cuts typically result in
improvements by factors of 10 to 100.

\subsection{Optics simulation}

\begin{figure}[htb]
 \ifwithtwocolumns
   \resizebox{\hsize}{!}{\includegraphics{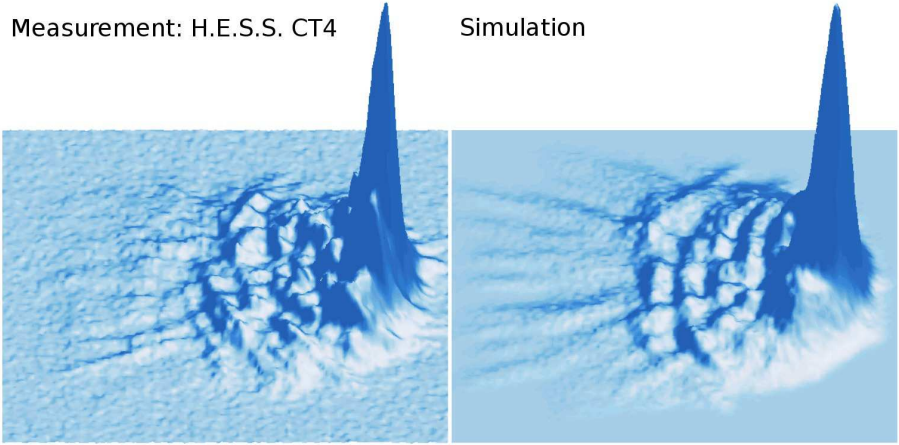}}
 \else
   \resizebox{12cm}{!}{\includegraphics{star-psf}}
 \fi
 \caption{Measured (left) and simulated (right) distribution of star light
   imaged onto the lid of a H.E.S.S. camera at an off-axis angle of
   2.3$^{\circ}$, adapted from {\citep{Cornils-PhD}}. The fields shown are
   about 10.5 cm (0.40$^{\circ}$) wide.\label{fig-star-shape}}
\end{figure}

All IACTs to date have segmented reflectors made with mirror tiles of
spherical type. The optical ray-tracing in \tmtexttt{sim\_telarray} fully
reflects this current usage, with spherical mirror tiles of round, hexagonal,
or square shape. The optical quality of the tiles can be adjusted to
laboratory measurement, including variations in focal length or non-perfect
surfaces. Also the mirror adjustment, typically performed with star light
falling onto the lid of the camera, will have a limited accuracy and is taken
into account when every individual mirror tile is configured at program
start-up. How well this approach can represent real telescopes is demonstrated
in the case of the H.E.S.S.~I reflectors, based entirely on mirror quality lab
measurements and the measured accuracy in the mirror alignment procedure
{\citep{HESS-optics-1}}{\citep{HESS-optics-2}}, without any free parameters.
Not only are the general characteristics of the optical PSF reproduced but
also the detailed shapes of star light imaged onto the camera lid
\citep{Cornils-PhD} (see Figure~\ref{fig-star-shape}),
at least for the zenith angle range where the
alignment was carried out. The zenith-angle dependence of the alignment
quality, due to dish deformations, can be set as well in order to match
measurements or separate finite elements calculations.

The optics is defined by the overall focal length $f_{\tmop{tel}}$, a table of
mirror positions $(x_i, y_i)$ perpendicular to the optical axis, shapes, and
sizes as well optional $z_i$ positions and mirror tile focal lengths $f_i$. If
the $z_i$ positions are omitted, mirrors will be positioned following variants
of either a Davies-Cotton or parabolic dish shape, taking mechanical
accuracies into account. Mirror tiles can have either all the same focal
lengths, specific $f_i$ from the table, or automatically optimized $f_i$
derived from the desired dish shape and $(x_i, y_i)$.

In the case of a parabolic dish, the optimal mirror tiles would clearly be
paraboloid segments, while all IACTS so far have been equipped with spherical
tiles. The optimum focal lengths of spherical tiles on a parabolic dish can be
obtained by calculating the principal (maximum and minimum) radii of
curvature of the paraboloid at the position of the mirror segment. Half the
geometric mean, i.e.\ 0.5 $(r_{\min} r_{\max})^{1 / 2}$, is used for automatically
optimized $f_i$ but the simple distance of the mirror tile to the system focus
would be reasonable as well, as the two lead to almost the same result.
Practical aspects of the mirror production for a real IACT may dictate to
build mirror tiles of only a few different focal lengths or even a single
focal length, within specifications. The optical ray-tracing of
\tmtexttt{sim\_telarray} is well suited for comparing the optical point-spread
functions (PSF) of different designs and optimize the design.

\begin{figure*}
\ifwithtwocolumns
 \resizebox{0.33\hsize}{!}{\includegraphics{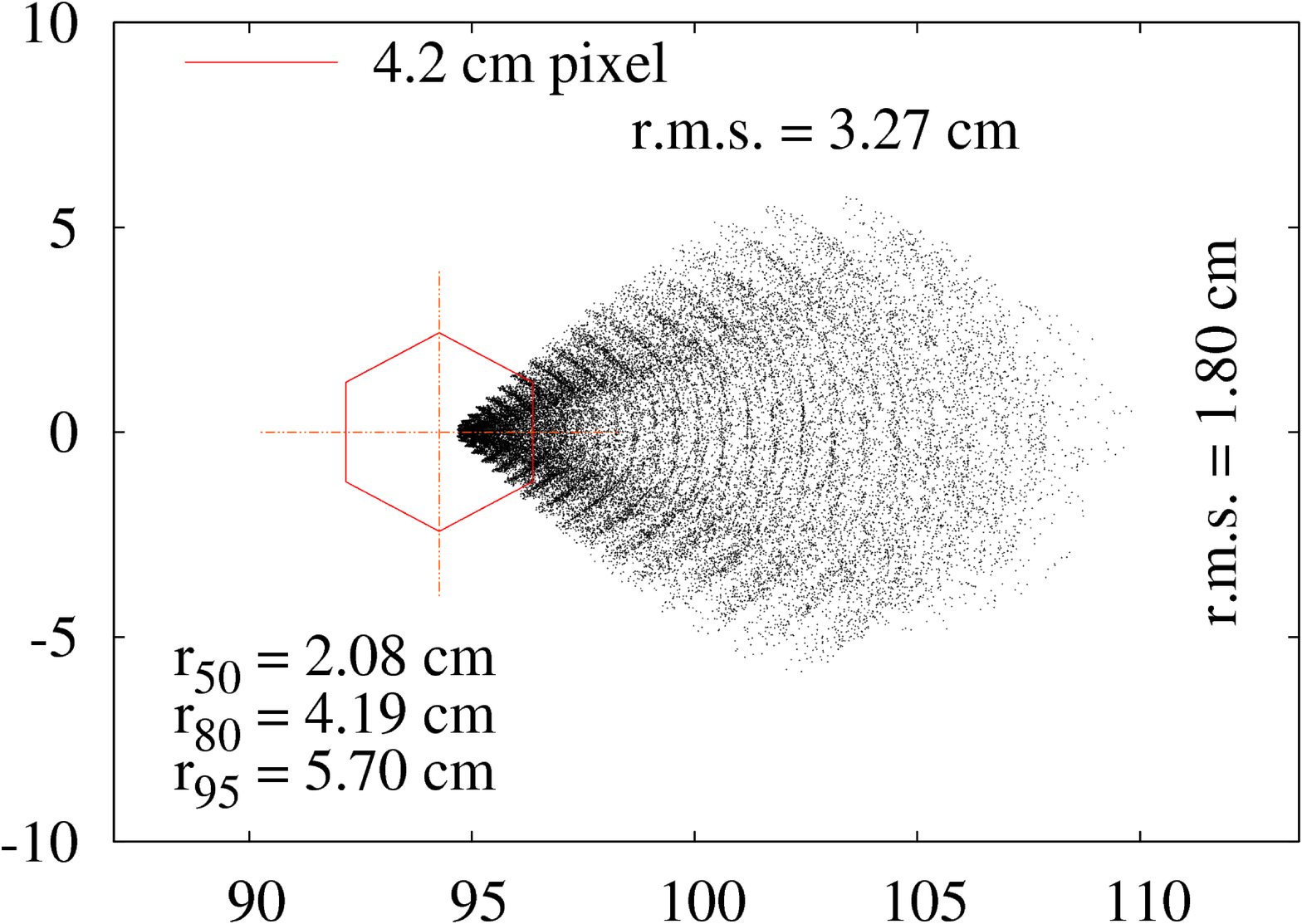}}
 \resizebox{0.33\hsize}{!}{\includegraphics{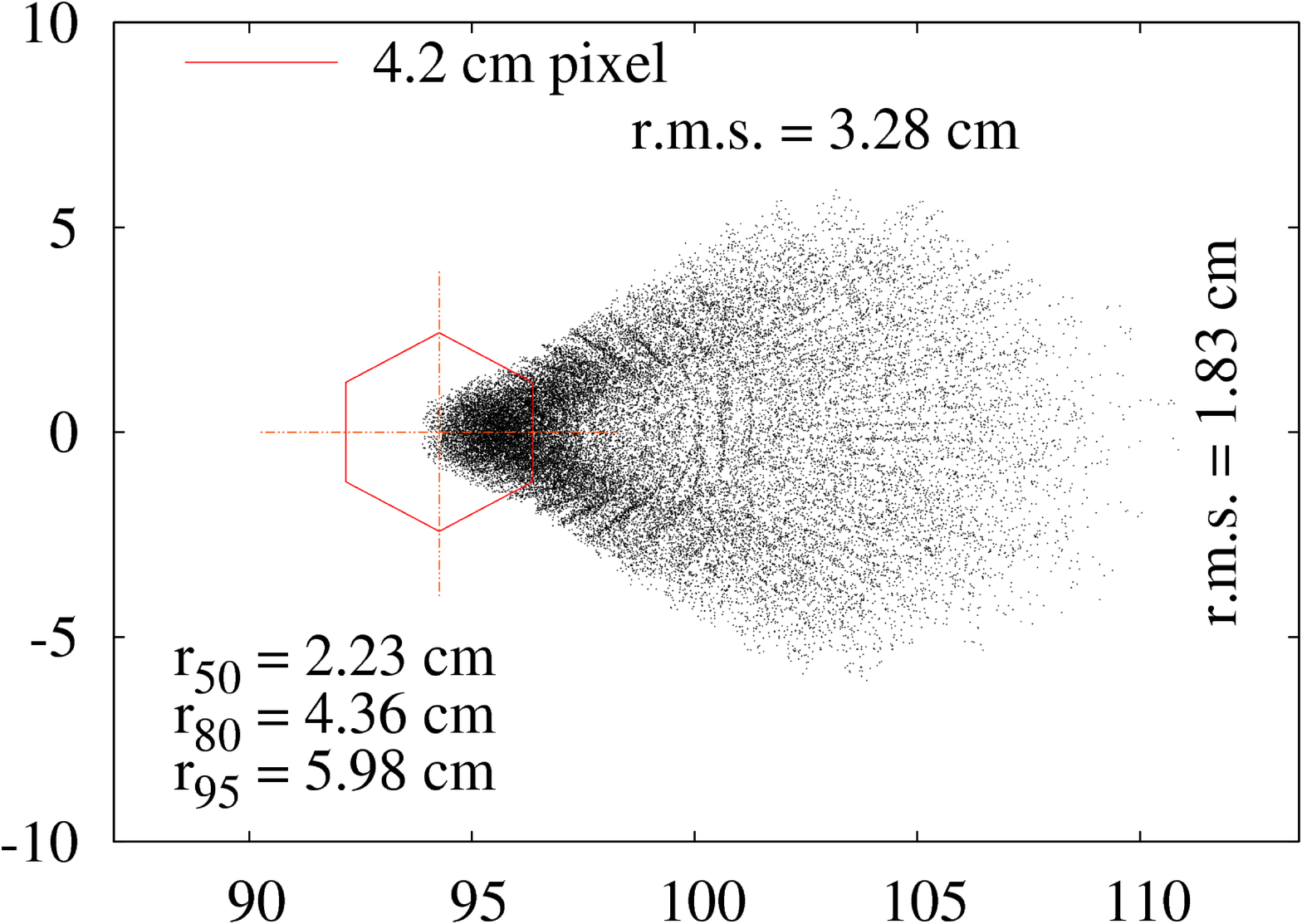}}
 \resizebox{0.33\hsize}{!}{\includegraphics{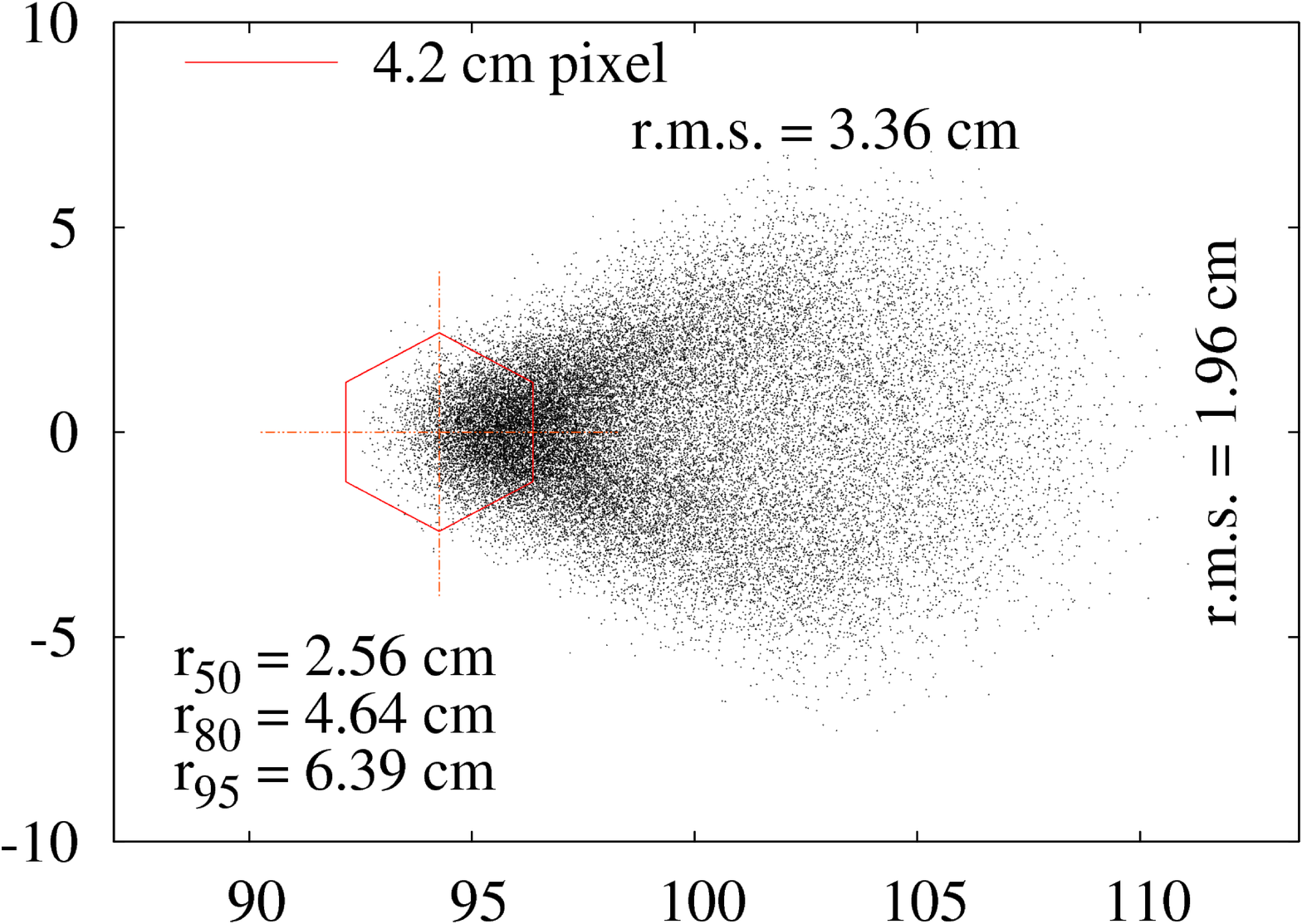}}
\else
 \resizebox{0.46\hsize}{!}{\includegraphics{psf-hess2-a}} \\
 \resizebox{0.46\hsize}{!}{\includegraphics{psf-hess2-b}} \\
 \resizebox{0.46\hsize}{!}{\includegraphics{psf-hess2-c}}
\fi
 \caption{Study of the point spread function of (Cherenkov) light from 
   a point source at 10 km distance and 1.5$^\circ$ off-axis
   imaged to the pixel entrance plane in a H.E.S.S.-2 type telescope ($f=36$~m, 
   coordinates in units of centimetres).
   The worst case off-axis orientation, along the larger vertical dish diameter,
   is assumed. 
   \ifwithtwocolumns Left: \else Top: \fi
   ideal mirror alignment and individually adapted mirror
   tiles. Middle: all mirror tiles of identical focal length. 
   \ifwithtwocolumns Right: \else Bottom: \fi
   in addition with realistic mirror alignment accuracy. R.m.s.\ widths of the light
   distributions along radial and transverse directions are indicated in
   the figures, as well as 50\%, 80\%, and 95\% containment radii.
   Note the extremely fine pixels of $0.067^\circ$.
   \label{fig-hess2-psf}}
\end{figure*}

An example study for the H.E.S.S.-2 telescope now under construction is
illustrated in Figure~\ref{fig-hess2-psf}, demonstrating that the
optical PSF is only marginally deteriorated by choosing a single
focal length for all mirror tiles.

Since the propagation time of light is included in the optics
simulation, the time spread due to non-isochronous mirror configurations
is automatically reproduced.
While a parabolic dish guarantees the shortest possible Cherenkov pulse
durations, different dish shapes may result in a still acceptable time spread
while improving the off-axis PSF. The \tmtexttt{sim\_telarray} program
therefore allows for variations of the assumed dish shape with respect to
either a Davies-Cotton or parabolic base design. Since the most important
aspect here is the radius of curvature of the dish this is largely equivalent
also to elliptical intermediate dish shapes.

Mirror tiles can be of hexagonal, square, or circular shape. Since the test
for intersection with every mirror tile could be very time consuming in the
simulations, a grid scheme somewhat similar to that in the CORSIKA IACT option
is applied, this time along the dish onto which the mirror tiles are mounted.
A mirror may be associated with one or more grid cells. By intersecting the
path of the incoming photon with the dish one grid cell gets selected and the
detailed intersection has to be evaluated only for the few mirror tiles
associated with that cell. Every mirror tile is once set up with random
misalignments and misplacements according to realistic specifications. Its
actual focal length may randomly vary around the defined or optimized value.
The surface is assumed to exhibit some random roughness resulting in a small
smearing of the reflected rays.

As an optional feature, not normally used for performance reasons, the support
structure for the camera can be included in the ray-tracing, approximated by
an arbitrary number of cylinders. This way, the {\tmem{shadowing}} by the
camera support structure and by the camera itself and its open lid can be
evaluated in detail, taking both incoming and reflected photons into account
{\citep{HESS-optics-1}}.
For normal production, the shadowing by the support structure is parametrized
as a function of the angle w.r.t.\ the optical axis and only the shadowing by
the camera is explicitly included.

In order to make such optical studies possible, and also for realistic tests
of the impact of a non-uniform NSB on the shower reconstruction,
\tmtexttt{sim\_telarray} allows to set up additional light sources, normally
at infinite distance (stars) but optionally at finite distance to evaluate the
imaging of specific positions along a shower axis. For the optical studies the
listing of reflected photons can be written to a file. For the shower
simulations, the additional stars result in an enhanced noise in the
illuminated pixels, for efficiency reasons determined once at program start-up.
The average or {\em direct current} (DC) of stars in the
photo tubes -- which gets decoupled by high-pass filters in real detectors --
is subtracted and not present in the signals.

With mirror tiles aligned to focus star light on the closed lid of the camera,
the Cherenkov light from the showers is focused behind the lid, at an offset
\begin{eqnarray*}
  d = (f^{- 1} - D^{- 1})^{- 1} - f
\end{eqnarray*}
for light emitted at a distance $D$ from the telescope of focal length $f$.
See {\citep{Hofmann-focusing}} on possible strategies for optimal focusing on
showers.

As a final step of the optics simulation, the angular acceptance of the pixels
is taken into account since pixels are typically equipped by Winston cones or
such to avoid gaps between neighbouring pixels as far as possible. For
practical reasons, this is made part of the photon detection efficiency. As
far as the ray-tracing is concerned, it is sufficient to find out which (if
any) pixel is hit at the entrance of its Winston cone or similar and
optionally if the photo-cathode at the base of the Winston cone is hit
directly or not. The shapes of pixel entrance and unobstructed photo-cathode
can be hexagonal, square, or circular. A more detailed set-up, activated in
the camera configuration file, may include
angular acceptance tables, averaged over the pixel area, as obtained from
measurements or detailed ray-tracing of the light collectors. Whenever the
relevant data is available, the more detailed mode is usually preferred.

\subsection{Photon detection}

The photon detection probability $\varepsilon$ basically depends on the
atmospheric transmission $T$, shadowing factor $S$ by the camera and its support
structure, the mirror reflectivity $R$, the efficiency $E$ of the light
concentrator (optionally whether a photon hits a pixel directly onto the
photo-cathode or via reflection on the light concentrator), on the quantum
efficiency $Q$ of the photo-cathode and finally the collection efficiency $C$
inside the PMT. Most of them are functions of wavelength:
\begin{eqnarray*}
  \varepsilon (\lambda) & = & T (\lambda) S R (\lambda) E (\lambda) Q
  (\lambda) C (\lambda).
\end{eqnarray*}

\begin{figure}[htb]
\ifwithtwocolumns
  \hc{\resizebox{0.9\hsize}{!}{\includegraphics{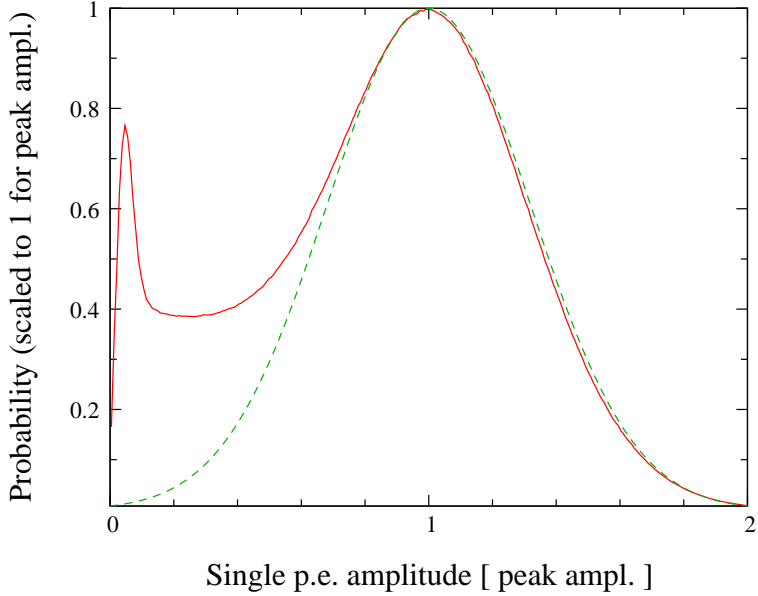}}}
\else
  \resizebox{10cm}{!}{\includegraphics{spe}}
\fi
  \caption{\label{fig:spe}Single-p.e. amplitude distribution at the PMT anode
  for the two ways to take a 85\% PMT collection efficiency into account.
  Solid line: collection efficiency included in the amplitude distribution
  (the mean amplitude is 85\% of the amplitude in the peak of the
  distribution). The distribution here is based on a detailed simulation of
  the multiplication process in PMTs with parameters fitted to match low-noise
  measurements between 0.1 and 3 times the peak amplitude. Dashed line:
  collection efficiency is included with the quantum efficiency and a normal
  distribution is assumed.}
\end{figure}

\begin{figure}[htb]
\ifwithtwocolumns
  \resizebox{\hsize}{!}{\includegraphics{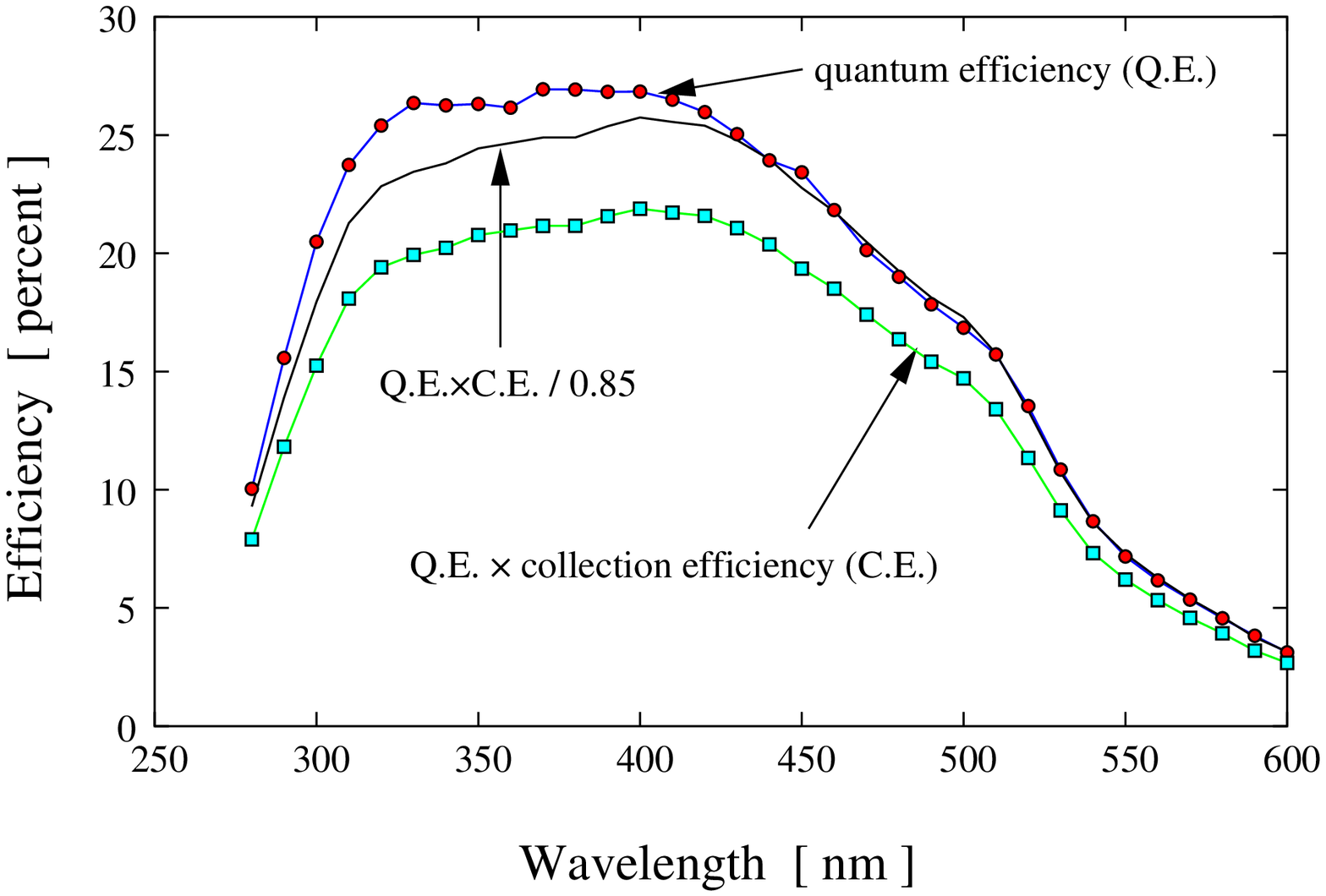}}
\else
  \resizebox{10cm}{!}{\includegraphics{qe_x_ce}}
\fi
  \caption{Quantum efficiency (upper curve), quantum efficiency times
  collection efficiency (lower curve), and the latter divided by the mean
  collection efficiency of 85\% (middle curve) as evaluated for a sample of
  PMTs for the H.E.S.S. experiment {\citep{Koch-Kohnle}}. The
  \tmtexttt{sim\_telarray} program can be either configured to include the
  mean collection efficiency in the single-p.e. amplitude distribution
  (together with the middle curve for the effective quantum efficiency) or to
  include it with the quantum efficiency (using the lower
  curve).\label{fig:qe_x_ce}}
\end{figure}

While fluctuations in the reflectivity of individual mirror tiles are smoothed
out, the fluctuations in quantum efficiency have to be taken into account.
Such fluctuations in quantum efficiency from PMT to PMT are usually obtained by measuring the 
photocathode currents of all PMTs induced by a calibrated light source, 
e.g.\ the CB (Corning Blue) value provided by the manufacturer.
The collection efficiency of typically 80 to 90\% is the probability that a
photo-electron actually hits the first dynode of a PMT and is effectively
multiplied rather than elastically scattered. It can be accounted for in two
ways. Either by including it in the detection probability as above and the
pulse amplitude distribution at the anode only accounting for photo-electrons
multiplied at the first dynode. Or by including the non-amplified
photo-electrons in the amplitude distribution of single photo-electrons 
(\textit{single p.e.}),
see Figure \ref{fig:spe}. For that purpose, the single-p.e. amplitude is randomly
drawn according to a table rather than from a normal distribution. The small
wavelength dependence of the collection efficiency is a consequence
of the photo-electric effect. It may be accounted for by adjusting the
quantum efficiency according to
\begin{eqnarray*}
  \tilde{Q} (\lambda) & = & Q (\lambda) C (\lambda) / \bar{C},
\end{eqnarray*}
where $\bar{C}$ is the average or effective collection efficiency. In Figure
\ref{fig:qe_x_ce} this is illustrated for PMTs evaluated for H.E.S.S.

For a realistic detector response, both the quantum efficiency and the voltage
required to achieve the desired gain can vary from PMT to PMT. For simplicity,
the same scaling factor is applied to the quantum efficiency at all
wavelengths. Variations of the voltage $U$ result in variations of the PMT
transit time which is proportional to $1 / \sqrt{U}$. The simulation also
includes a transit time jitter for individual photo-electrons.

The pixels are permanently exposed to NSB at a level configurable for every
pixel (including light of additional bright stars), in units of photo-electrons
per nanosecond. For existing detectors, this number is obtained from 
the image noise. An utility program is provided for evaluating the impact of 
different quantum efficiency curves, different mirror reflectivity etc. on the 
NSB rate, assuming the NSB spectrum measured by Benn and Ellison \citep{Benn1998}.
In real detectors, the DC (direct current) contributions 
get decoupled by high-pass filters.
In the simulation, the mean amplitude of NSB signals is therefore subtracted 
from the signal baseline. For every random NSB photo-electron, the corresponding
pulse shape with randomized amplitude is added on top of that baseline.

\begin{figure}[htb]
\ifwithtwocolumns
  \resizebox{\hsize}{!}{\includegraphics{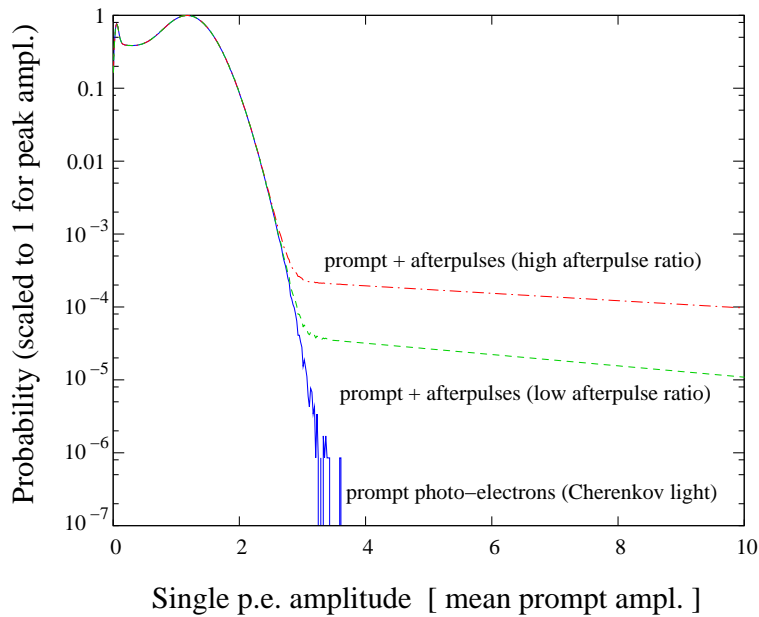}}
\else
  \resizebox{10cm}{!}{\includegraphics{eval_spe}}
\fi
  \caption{Amplitude distribution for prompt single-p.e. signals
  and after including the afterpulses (fits to two types of PMTs, as
  compared in Figure~\ref{fig:afterpulse}).
   \label{fig:afterpulse-spe}}
\end{figure}

\begin{figure}[htb]
\ifwithtwocolumns
  \resizebox{\hsize}{!}{\includegraphics{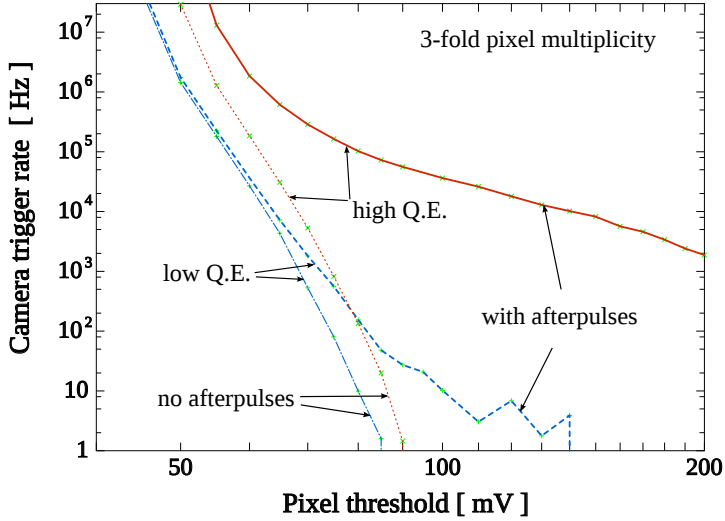}}
\else
  \resizebox{10cm}{!}{\includegraphics{nd5-3fold}}
\fi
  \caption{Example evaluation of PMT specifications for use in a future
  Cherenkov telescope. One PMT type has a slightly higher quantum efficiency
  but also a five times higher afterpulse ratio than the other 
  (see Figure \ref{fig:afterpulse-spe}). Simulations
  assume a mean single-p.e. amplitude of 23 mV for both and the same night-sky
  background resulting in random camera triggers (without any Cherenkov
  light). For the thin dotted and dash-dotted lines the afterpulsing is
  ignored, resulting in only a small increase in pixel comparator thresholds
  required with high Q.E. PMTs to achieve the same trigger rate as with low
  Q.E. PMTs. When afterpulsing is included (thick solid and dashed lines), the
  higher Q.E. PMT turns out to be unusable, at least with 3-fold pixel
  coincidences, because the required threshold would be at least three times
  as large as for the low Q.E. PMT type.\label{fig:afterpulse}}
\end{figure}

Another important NSB aspect taken into account is the
afterpulsing typically induced by ions in the PMT hitting the photo-cathode.
These afterpulses typically come hundreds of nanoseconds after the original
electron cascade by which the atom was ionized. For Cherenkov photons, the
afterpulses will therefore always be well outside the readout window. NSB
photons registered hundreds of nanoseconds earlier though can be well
responsible for afterpulses within the time window relevant for the trigger
and signal readout. The amplitude distribution of afterpulses extends to
much larger amplitudes than that of normal signals (see Figure~\ref{fig:afterpulse-spe}).
This fact is taken into account by having different 
single-p.e. amplitude distributions for Cherenkov light and for NSB, with the
afterpulse amplitude distribution added for the NSB signals only. How
important proper accounting for afterpulses can be is demonstrated in Fig.
\ref{fig:afterpulse}.

\subsection{Electronic signals and trigger}

After the photons are ray-traced to the PMT pixels and there may be a chance
of triggering of the telescope, based on the number of photo-electrons,
detailed simulation of the electronics response is started. Several different
pulse shapes are involved. These begin with the pulse shape for a single p.e.
at the input of a comparator or discriminator for each pixel. Similar but
different pulse shapes apply for each channel in the digitization process. Two
ways of having more than one channel per pixel are supported here: different
pre-amplifier gains for an enhanced dynamic range (as used with H.E.S.S.) and
multiple, phase-shifted FADC modules for achieving a high sampling rate (as
used with HEGRA).

For the analogue signal at the comparator or discriminator input, an internal
sampling short compared to actual pulse durations is required. Typically, a
250 ps time interval is used for pulses of a few nanoseconds. Where the camera
digital electronics should provide only an integrated signal, a sufficiently
fine-grained internal sampling is used in the first place and the signal is
added up afterwards. Otherwise the actual sampling of the read-out system is
used and the common phase of the digitization (across the whole camera) is
randomized with respect to the photon arrival times. For each photo-electron,
the pulse shapes are scaled by the random single-p.e. response and shifted
according to the photon arrival time plus PMT transit time including a random 
jitter. All single-p.e. signals from the Cherenkov light as well as the 
NSB photons are added up. 
For pixels exposed to too much star light, limits are available
when the pixels are disabled in the trigger and when their HV would be
turned off to avoid damage.

\begin{figure}[htb]
\ifwithtwocolumns
  \resizebox{\hsize}{!}{\includegraphics{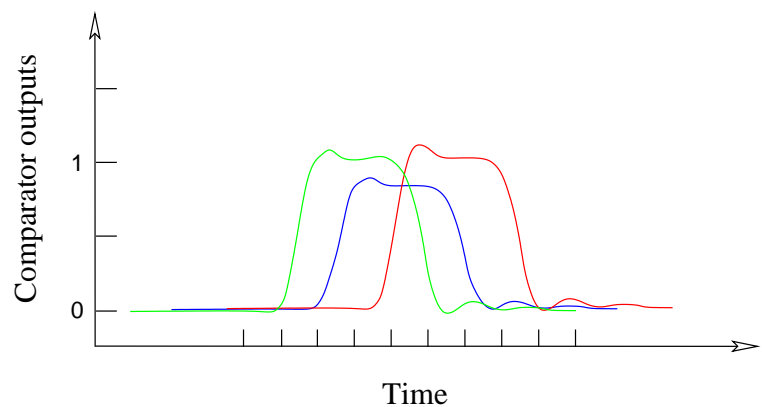}}
\else
  \resizebox{10cm}{!}{\includegraphics{comp-sig-scheme}}
\fi
  \caption{\label{fig:comp-pulse-scheme}Three comparator or discriminator
  output signals partially overlapping in time may or may not be sufficient
  for a telescope trigger set to a 3-fold multiplicity threshold. Detailed
  treatment of trigger switching is required.}
\end{figure}

\begin{figure}[htb]
\ifwithtwocolumns
  \resizebox{\hsize}{!}{\includegraphics{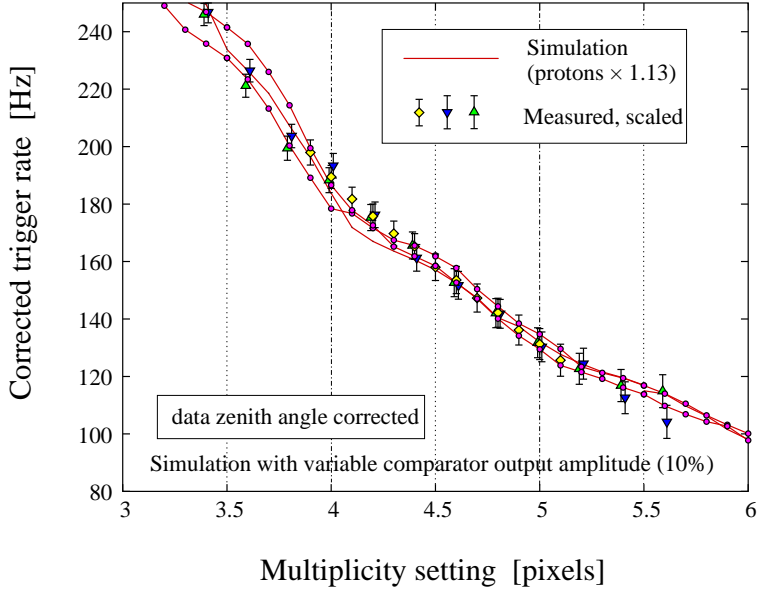}}
\else
  \resizebox{10cm}{!}{\includegraphics{trgsim5}}
\fi
  \caption{\label{fig:hess-mult-rate}Telescope trigger rate as a function of
  the threshold level corresponding to the pixel multiplicity. Data from the
  first H.E.S.S. telescope {\citep{HESS-trigger}}, taken in different nights
  and corrected for zenith angle dependence. Simulations including only proton
  showers were scaled up in absolute rate by a common factor to match the
  data (no absolute calibration being available at that time).
  The telescope triggers when the (analogue) sum of the pixel comparator
  outputs in any {\em sector} of 64 pixels can switch the sector comparator
  with programmable threshold. That threshold is shown here in units of
  the average pixel comparator output, i.e.\ effectively the number of
  pixels fired at the same time.
  Idealized simulations would show steps at integer multiplicity values,
  while the realistic simulations show only a very modest flattening between
  integer values. Three sets of simulations were carried out, with random
  variations of quantum efficiencies, comparator output levels, etc., as
  indicated by manufacturer specifications.}
\end{figure}

For a realistic response of the trigger system, the switching behaviour of the
comparators or discriminators and the rise and fall times of their output
signal -- typically on the order of a nanosecond -- has to be taken into
account, as illustrated in Figure \ref{fig:comp-pulse-scheme}. And they will
not switch instantly when the signal exceeds the predefined threshold. This
can be either achieved by demanding a minimum time-over-threshold or a
switching charge (minimum area between pulse shape and threshold level). The
\tmtexttt{sim\_telarray} program even takes possible variations in the output
amplitude from device to device -- 10\% being not uncommon -- into account.
Since the telescope trigger decision is based (at least effectively) on a
second comparator or discriminator decision applied to the sum of pixel logic
outputs, this has testable consequences. A fully digital trigger decision
would correspond to zero rise and fall times as well as identical
amplitudes of all pixel logic outputs. In reality, the signals are so
short that the pulse shape of pixel logic outputs should not be ignored.  
As an example, the -- originally
unexpected -- smooth change in trigger rate of the first H.E.S.S. telescope as
a function of the multiplicity threshold could be readily explained by the
available simulations, with comparator properties according to manufacturer
specifications. This is illustrated in Figure \ref{fig:hess-mult-rate}.

The telescope trigger logic, i.e.\ comparator or discriminator outputs from
which pixels can be combined to form a telescope trigger, is different for
each IACT array. In HEGRA a {\tmem{next-neighbour}} logic was used and in
H.E.S.S. a {\tmem{sector}} logic based on $8 \times 8$ pixel sectors, while
Smart Pixel {\citep{SmartPixel}} and other designs differ again. The
\tmtexttt{sim\_telarray} program aims to be fully flexible in this area and
its configuration files include lists of all pixel combinations from which a
telescope trigger can be formed via a (multiplicity) threshold in the sum of
corresponding comparator/discriminator outputs.

The last step in the trigger decision of an IACT array will be the system
trigger, usually as a requirement of at least two telescope triggers within a
short time window, after correcting for the expected time delay as a function
of the system viewing direction and the telescope positions. The required
width of the window -- typically several 10 ns -- depends not only on the
fields-of-view of the telescopes but also on the stability of the transmission
lines (cable or optical fibres) from the telescopes to the central or
distributed system trigger logic. Since random system trigger rates are
typically very low and the width of the system trigger time window therefore
usually chosen wide enough, any changes in transmission delays can be ignored
by \tmtexttt{sim\_telarray}. Since the delay times are compensated
automatically in \tmtexttt{sim\_telarray}, the telescope multiplicity and the
width of the coincidence time window are usually the only parameters needed.
Optionally, \tmtexttt{sim\_telarray} allows read-out of non-triggered
telescopes, as it happened to be the case for the HEGRA IACT array.

\subsection{Telescope raw data and other output data}

The main output data from \tmtexttt{sim\_telarray} should correspond to what
would be recorded from real telescopes ({\tmem{raw data}}) as closely as
possible. This can be full {\tmem{sample-mode}} digitized pulse shapes for
every pixel or the signal sum over the given integration window. Different
types of zero-suppression are possible, like suppression of low-gain channels
when the amplitude in corresponding high-gain channels is small or full
suppression of pixels with small amplitudes.

The underlying data format is based on the {\tmem{eventio}}
machine-independent format, like the CORSIKA IACT output itself but using more
and different data block types. Strictly simulation-related data, like the
nature, direction, and true energy of the primary particle or the actual
displacements of telescope arrays with respect to the nominal core position,
are kept separate from the raw data, i.e.\ in separate data blocks but in the
same data stream or file, and can be optionally omitted or filtered out later
for realistic data challenges. The data format is largely independent of that
used by past and current IACT arrays like HEGRA and H.E.S.S. but conversion to
experiment-specific formats is generally straight-forward and provided through
utility programs.

Data that is needed for calibration, like conversion from ADC counts to
photo-electron numbers, laser/LED flat-fielding, muon-ring efficiency etc. can
be derived from full simulations of the various types of calibration runs. The
more basic numbers like pedestals, photo-electron conversion factors, 
and flat-fielding coefficients are also directly provided in the output data.
For the pedestals, which may depend on a varying temperature and other
things in a real detector, a random `measurement' error with configurable
r.m.s.\ is applied to the reported value.

\subsection{Built-in shower reconstruction and event visualization}

\begin{figure}[htb]
\ifwithtwocolumns
  \resizebox{\hsize}{!}{\includegraphics{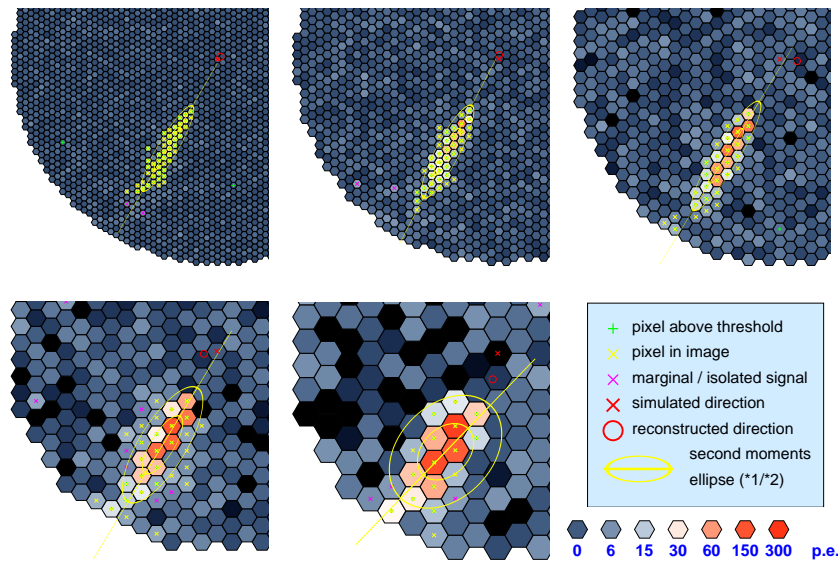}}
\else
  \resizebox{11cm}{!}{\includegraphics{pixel_size_image}}
\fi
  \caption{Example of the \tmtexttt{sim\_telarray} event visualisation,
  showing the same shower (a $\gamma$-shower of 460 GeV energy at 190~m core
  distance) in telescopes with different pixel sizes of 0.07 to
  0.28$^{\circ}$, in a 9-telescope array evaluated for the CTA project
  {\citep{CTA-MC-ICRC}}. For clarity, only part of the field-of-view of one
  and the same of the nine telescopes is shown for each pixel size. The second
  moments ellipses are shown at one and two times their actual size, with the
  major axes drawn up to four times their actual length. Reconstructed and
  actual shower directions are indicated by the small circles and crosses near
  the camera centres, respectively.\label{fig:event-display}}
\end{figure}

The \tmtexttt{sim\_telarray} program includes a basic geometrical shower
reconstruction, based on second moments (Hillas parameters) and stereoscopic
reconstruction methods originally developed for HEGRA. The shower direction is
defined by the weighted average of the pair-wise intersection points of Hillas
second-moments major axes in every two-telescope combination with two
well-defined images. Only those reconstruction steps are carried out which do
not require any look-up of other simulation data. Energy lookup and
gamma-hadron-discrimination are thus precluded here. Instead, the
reconstruction is oriented towards immediate visualisation of the data. A
dedicated Postscript-generator for very compact plot output was developed for
that purpose. See Figure \ref{fig:event-display} for an example.

\section{Conclusions and outlook}

The CORSIKA program with the IACT/ATMO extensions aims to provide the most detailed
simulation of air showers and the production of atmospheric Cherenkov light,
both for imaging and non-imaging detectors. For efficiency reasons not every
possible detail is normally used, like the wavelength dependence of the index
of refraction, but is available on demand. A variety of options are at hand to
adapt the level of detail and performance as needed for specific applications.
Parts of the photon detection probability can be applied within CORSIKA or can
be provided by the second simulation stage, the detector simulation. The
machine- and compiler-independent {\tmem{eventio}} data format allows for a
variety of schemes. The IACT extension, normally but not necessarily using
this flexible data format, allows treatment of the largest telescope or
detector arrays envisaged for coming generations of atmospheric Cherenkov
experiments or observatories. It is part of the standard CORSIKA distribution
for several years now.

For the second stage of the simulation process, a dedicated program for
imaging telescopes with tessellated primary mirrors, termed
\tmtexttt{sim\_telarray}, aims to provide the greatest level of detail in the
simulation of all optical and electronics components involved in the
measurement process. At the same time, \tmtexttt{sim\_telarray} is highly
efficient. The {\tmem{eventio}} data from the CORSIKA IACT option can be piped
concurrently into several \tmtexttt{sim\_telarray} processes, each with a
different configuration, while avoiding I/O bottlenecks on large computer
clusters.

The simulation results from \tmtexttt{sim\_telarray} can be either
written directly into an experiment-specific format, as initially done with
the HEGRA IACT system, or into a more generic, {\tmem{eventio}}-based format.
The resulting data can be analysed directly from the latter format or
converted into experiment-specific formats, e.g.\ the H.E.S.S. data format, and
then analysed with the same software as measured data.

A wide range of cross checks with measured data -- mainly by H.E.S.S. -- 
confirms that CORSIKA and \tmtexttt{sim\_telarray} can reproduce the
data extremely well, for newly commissioned telescopes just based on
pre-production lab measurements of mirror qualities, quantum efficiencies
and so on. Examples of that include the optical PSF {\citep{HESS-optics-2}},
the intensity, ring radius, and ring width of muon ring images {\citep{Bolz-PhD}},
image shapes of gamma and background showers 
as well as the reconstructed shower PSF for point-like gamma-ray sources
{\citep{HESS-Crab,HESS-PKS2155}}, as well as the trigger rates
{\citep{HESS-trigger}}. After commissioning, H.E.S.S. simulations only
had to be adapted to the slowly degrading optical throughput of the telescopes.

Future developments in \tmtexttt{sim\_telarray} may include non-spherical
mirror tiles or secondary mirrors, and extensions to the photon detection
code, as detectors with novel properties come into operation. Since the aim is
to provide data like a real instrument, more capabilities of future read-out
systems, like advanced post-processing in FPGAs which may include peak finding
and shape analysis, are among the most recent additions.

The CORSIKA program together with the IACT/ATMO package is available from
Forschungszentrum Karlsruhe. For details see \\
\texttt{http://www-ik.fzk.de/corsika/}. 

The \tmtexttt{sim\_telarray} program
is available on request from the author. A release under
an open source licence is foreseen.


\section*{Acknowledgements}

It is a pleasure to thank D. Heck and T. Pierog for the development and the
continued maintenance of CORSIKA in the first place but also for the seamless
integration of Cherenkov emission and the IACT/ATMO extensions into CORSIKA
and its build process. Quite a number of other people have made important
contributions to the variety of options available for Cherenkov light emission
in CORSIKA. The \tmtexttt{sim\_telarray} program has seen extensive testing
with data from the HEGRA and H.E.S.S. experiments. A large effort by many
colleagues went into the data by these experiments as a whole and also into
the many lab and field measurements which were useful for the various
configuration files, from quantum efficiencies and mirror reflectivities over
pulse shapes, detailed single-p.e. measurements to mirror point spread
functions and mirror alignment etc.


\bibliographystyle{elsart-num}
\bibliography{mcpaper}

\end{document}